\documentclass[onecolumn]{aa}
\usepackage{graphicx}
\usepackage{amsmath}

\newfont{\myfont}{cmmib10}
\newcommand{\bkappa}{\hbox{\myfont \symbol{20} }}

\newcommand{\btheta}{\hbox{\myfont \symbol{18} }}
\newfont{\myfontsmall}{cmmib8}
\newcommand{\bkappasm}{\hbox{\myfont \symbol{20} }}

\newcommand{\bthetasm}{\hbox{\myfontsmall \symbol{18} }}

\begin{document}

\title{Scintillation-induced Variability in Radio Absorption Spectra against Extragalactic Sources}

\authorrunning{J.-P. Macquart}
\titlerunning{Spectral variability measured against extragalactic sources}

\author{Jean-Pierre Macquart}

   \offprints{J.-P. Macquart}
   \institute{Kapteyn Astronomical Institute, University of Groningen, The Netherlands\\
                \email{jpm@astro.rug.nl}
             }


\abstract{Spectral features absorbed against some radio quasars exhibit $\sim 50\,$mJy variations, with the lines varying both relative to the continuum and, when several lines are present, even relative to one another.  We point out that such variability can be expected as a consequence of refractive scintillation caused by the interstellar medium of our Galaxy.  Scintillation can cause independent variations between closely-spaced spectral lines, and can even alter the line profile.  The background source need not be compact to exhibit spectral variability.  The variability can be used to infer the parsec to sub-parsec scale structure of the intervening absorbing material.  We discuss the importance of scintillation relative to other possible origins of spectral variability.  The present theory is applied to account for the variations observed in the HI-absorbed quasar PKS\,1127$-$145.   
}

   \maketitle
%

\section{Introduction}
There are now two well-established cases of variability in HI-absorption spectra measured along lines of sight to high redshift quasars (Wolfe, Davis \& Briggs 1982; Kanekar \& Chengalur 2001).
Changes in the amplitudes of spectral features relative to one another and to the continuum are observed on time scales of days to weeks, and there is no evidence for shifts in the absorption frequencies with time.  Such rapid changes in absorption spectra imply the presence of structure in the intervening absorbing material on scales far below those attainable with conventional imaging techniques. 

The variations could be due to inhomogeneities in the absorbing source coupled with changes in the line of sight to the background quasar or effects due to interstellar scintillation. 
Spectral variability due to changes in the line of sight occurs if the background source moves at high angular speed with respect to the absorbing system.  The time scale of spectral variations is determined by the angular speed of the background source relative to the angular scale of fluctuations in the intervening absorbing system.

The interpretation of the variability in terms of interstellar scintillation is more troublesome. 
No changes in the absorption strength relative to the continuum are expected from the simplest scintillation model, in which radiation from a point-like source is absorbed by foreground material. The next simplest model consists of a background source composed of several components compact enough to scintillate, with different spectral features associated with each source component.  Variations in spectral features relative to the continuum are expected in this model, but appear unable to quantitatively account for the actual variations observed in these sources.  For PKS~1127$-$145, Kanekar \& Chengalur (2001) rule out a source model consisting of one point-like variable component and one component of constant flux density;  the model predicts that when the source exhibits the same continuum flux density it would also exhibit the same absorption spectrum, which is not observed.  Moreover, a source model composed of unrealistically many compact background components is required to account for the large number ($>4$) of independently varying spectral features that are actually observed.  An alternate scintillation model proposed to explain HI variability in pulsars (Gwinn 2001) can also be ruled out, as it requires the source to exhibit (diffractive) scintillation in the absorbing medium, which is not observed here.

However, existing models do not take into account effects related to the finite angular diameter of the background source.  The variability characteristics of the scintillation depend in detail on the source image structure.  This is important because, even for the most compact quasars, the presence of any sub-structure in the absorbing medium influences the apparent image of the absorbed source.  Thus the source image that scintillation responds to differs between the continuum and any spectral feature.   The implication of this simple point has not been appreciated in the discussion of spectral variability, and it has a profound influence on the variability of spectral features due to interstellar scintillation.  



In this paper we compute the effect of scintillation on absorption spectra, taking into account the effect of inhomogeneous structure imposed by the absorbing material across an extended background source.
The theory detailing the response of scintillation to image structure is presented in Sect. \ref{extragalactic}.  In Sect. \ref{applications} the theory is applied to compute the variability properties of some commonly encountered distributions of absorbing material.  The simplest model of absorption against a compact background source is used to account for the variations observed in PKS\,1127$-$145 in Sect. \ref{1127}.  The implications of scintillation for the interpretation of absorption spectra are discussed in Sect. \ref{implications}, and the conclusions are presented in Sect. \ref{conclusions}.  Throughout the text, we focus on systems observed in absorption.  However, the theory applies to equally to spectral lines observed in emission without a background source and we indicate how the results should be interpreted for such systems.

\section{Theory}\label{extragalactic}

Consider the radiation from a background radio source whose spectrum is altered by absorption lines due to a foreground system, as shown in Fig. 1.  As it enters our Galaxy, scattering by the diffuse ionized component of the interstellar medium (ISM) induces variability, or interstellar scintillation, in the intensity of the radiation.  The properties of the variability depend sensitively on the angular brightness distribution toward the background source, which is influenced by the angular distribution of the absorbing material (see Fig. 2).  In this section we describe how a given distribution of absorbing material determines the characteristics of the spectral variability due to scintillation.

We restrict the discussion to the effects of scattering in our Galaxy.  Scintillation in the ionized medium of the absorption system may, in principle, also induce spectral variability.  However, because the distance between the scattering and absorbing medium in this case is small, the apparent size of the HI absorbing structure is too large for it to be subject to the effects of scintillation.

\begin{figure*}
\centering
\includegraphics[width=15cm]{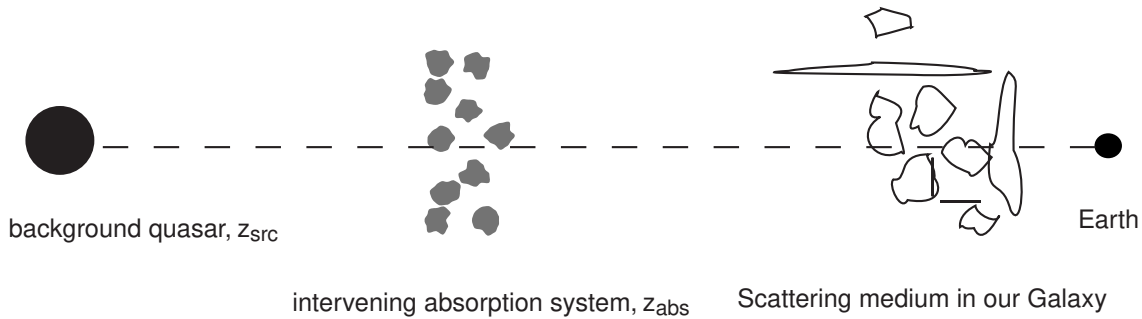} 
\caption{The geometry of the absorbing system.}
\label{HIimagesPicture}
\end{figure*}

\subsection{The scattering model}


For present purposes it is convenient to idealise the interstellar scattering material as being confined to a thin phase screen located at a distance $L$ from the observer.  This assumption is commonly used in astrophysical situations (Prokhorov et al. 1975; Rumsey 1975, Mercier 1962, Salpeter 1967), and a treatment incorporating the extended nature of the scattering medium introduces no additional effects relevant to the present discussion.

It is necessary to characterise the phase fluctuations responsible for the scattering.  Observations along many lines of the sight through the Galactic interstellar medium indicate that the power spectrum of density fluctuations in the ISM often follows a power law with index $-\beta$ close to the Kolmogorov value of $\beta=11/3$ (e.g. Armstrong, Rickett \& Spangler 1995).  The amplitude of the spectrum depends on the particular line of sight through the Galaxy, and is given by the scattering measure ${\rm SM} = \int dz \, C_N^2(z)$, where $C_N^2(z)$ denotes the level of turbulence as a function of distance along the ray path (see, e.g. Rickett 1977).  The information on phase fluctuations in the thin screen approximation is succinctly embodied in the phase structure function, 
$D_{\phi}({\bf r}) = \langle [ \phi({\bf r}'- {\bf r}) - \phi({\bf r}') ]^2 \rangle$, which takes the form
$D_\phi ({\bf r}) = \left( r/r_{\rm diff} \right)^{\beta-2}$ for $\beta < 4$ provided that $r$ exceeds the inner scale of the power law (Coles et al. 1987).  The diffractive length, $r_{\rm diff}$, is the scale at which the root mean square phase difference across the phase screen is unity, and it is defined formally in terms of the scattering measure as
\begin{eqnarray}
r_{\rm diff} = \left[ \frac{2^{\beta-2} (\beta-2)}{8 \pi^2 r_e^2 \lambda^2 \, {\rm SM}\, } \frac{\Gamma \left(1+\frac{\beta-2}{2} \right)}{\Gamma \left(1-\frac{\beta-2}{2} \right) } \right]^{1/(\beta-2)}.
\end{eqnarray}
For $\beta=11/3$ its numerical value is 
\begin{eqnarray}
r_{\rm diff} &=&1.525 \times 10^7 \, \left( \frac{{\rm SM}}{1.54 \times 10^{15}\,{\rm m^{-17/3}} } \right)^{-\frac{3}{5}} 
\, \, \left( \frac{\lambda}{1\,{\rm m} } \right)^{-\frac{6}{5}} \, {\rm m}, \label{rdiffDefn} 
\end{eqnarray}
where the scattering measure for a medium of thickness $\Delta L$ is 
\begin{eqnarray}
{\rm SM} = 1.54 \times 10^{15} \left( \frac{C_N^2}{10^{-4}\,{\rm m}^{-20/3}} \right) \, \left( \frac{\Delta L}{0.5\,{\rm kpc}}\right) \, {\rm m}^{-17/3}.
\end{eqnarray}

Armed with a description of the phase fluctuations on the phase screen, one can describe the power spectrum of intensity fluctuations across the observer's plane for a point source.  The source is assumed to be located at infinity -- an excellent approximation for the extragalactic sources under consideration here -- so that the wavefronts incident on the scattering medium may be considered planar.  For scintillation in the regime of weak scattering and refractive scintillation in the regime of strong scattering the power spectrum of intensity fluctuations is (Codona \& Frehlich 1987; Coles et al. 1987)
\begin{eqnarray}
P_{\rm scint}(\bkappa) = 8 \pi r_e^2 \lambda^2\, {\rm SM} \, \kappa^{-\beta} 
\sin^2 \left(\frac{\kappa^2 L}{2 k} \right)  \exp\left[ - D_{\phi}\left( \frac{\bkappa L}{k} \right)  \right]. 
\label{Codona11}
\end{eqnarray} 
Strong scattering occurs when $r_{\rm diff} < \sqrt{L/k}$ (e.g. Narayan 1992).  In this regime the spectrum in Eq. (\ref{Codona11}) is cut off by the exponential term at wavenumbers $\kappa=r_{\rm diff} \, (L/k)^{-1} \equiv r_{\rm ref}^{-1}$, where $r_{\rm ref}$ is the scale of refractive scintillations.  In the strong scattering regime the exponential term becomes important before the sine term oscillates rapidly with $\kappa$.  

Observations of pulsars and intra-day variable radio sources located off the Galactic plane indicate that objects viewed through the ISM at frequencies below $\sim 4$\,GHz are in the regime of strong scattering (Taylor \& Cordes 1993; Walker 1998; Kedziora-Chudczer et al. 1997, Dennett-Thorpe \& de Bruyn 1999).  Thus observations of HI absorption lines, for instance, are always in the regime of strong scattering.  The refractive scintillation that occurs in strong scattering is a broadband phenomenon, which is to say that a point source exhibits similar intensity variations over a bandwidth comparable to the observing frequency (i.e. over a range $\Delta \nu \sim \nu$).

Ultra-compact sources can also exhibit another manifestation of strong scattering, called diffractive scintillation.  The contribution of diffractive scintillation is omitted from the present discussion.  This narrowband phenomenon contributes at high wavenumbers $\kappa \sim r_{\rm diff}^{-1} \gg r_{\rm ref}^{-1}$.  This is neglected because terms that contribute at such high spatial frequencies are suppressed unless the source angular diameter is below $r_{\rm diff}/L$ which requires source sizes $\theta < 1\,\mu$as at $\nu \sim 1\,$GHz.   Not even the most compact extragalactic sources, including intra-day variable sources, appear to satisfy this requirement (e.g. Dennison \& Condon 1981; Kedziora-Chudczer, Macquart \& Jauncey 2001).

\subsection{Variability of an extended source}

In practice no extragalactic radio source can be regarded as point-like, and the power spectrum given by Eq.\,(\ref{Codona11}) needs to be modified to take into account both the angular structure of the extragalactic source and its foreground absorbing structure. 
The power spectrum of intensity fluctuations across the observer's plane is the product of the point source scintillation spectrum with the power spectrum of the source angular brightness distribution (Little \& Hewish 1966; Salpeter 1967; Rumsey 1975):
\begin{eqnarray}
P_I(\bkappa;\nu)&=& P_{\rm scint}(\bkappa) \left\vert V_{\rm app}\left( \frac{\bkappa L}{k} ; \nu\right) \right\vert^2. \label{PCentral}
\end{eqnarray}
The quantity $V_{\rm app}({\bf r};\nu)$ is the visibility of the source image at frequency $\nu$, or equivalently, wavenumber $k=2 \pi \nu/c$:
\begin{eqnarray}
V_{\rm app}({\bf r};\nu) = \int d^2 \btheta \, I_{\rm app}(\btheta;\nu) \exp \left( - i k \btheta \cdot {\bf r} \right), \label{Visdefn}
\end{eqnarray}   
The apparent image is determined by both the angular brightness distribution of the background source, $I_{\rm src}(\btheta;\nu)$, and the angular distribution of the opacity, $\tau(\btheta;\nu)$, due to  the intervening absorbing medium: $I_{\rm app}(\btheta;\nu)= I_{\rm src} (\btheta;\nu) \, e^{-\tau(\bthetasm;\nu)}$. 
The structure of the absorption material has a direct influence on the variability of its associated spectral line.  The scintillation properties of a source can change sharply with frequency when the angular distribution of the optical depth $\tau(\btheta;\nu)$ varies with frequency.   This is important if the absorption at each spectral line is associated with a distinct angular distribution of absorbing structure.  The background source itself need not be compact for the apparent image to contain compact features, since the absorption can also introduce compact features into the apparent image. Thus the background source itself need not be compact for spectral features to exhibit scintillation, as shown in Sect. \ref{extendedGauss} and Sect. \ref{extendedPower} below.

In practice an observer does not measure intensity fluctuations across a plane, but rather the temporal intensity fluctuations at a single point on this plane.  The two are related under the assumption that the phase fluctuations are frozen into the scattering screen (e.g. Rumsey 1975).  The spatial power spectrum of intensity fluctuations, $P_I(\bkappa)$, is related to the auto-correlation of the intensity fluctuations by a Fourier transform 
\begin{eqnarray}
C_I(t;\nu) = \langle \Delta I(t';\nu) \Delta I(t'+t;\nu) \rangle =  \int d^2 \bkappasm \, e^{i \bkappasm \cdot {\bf v}t} P_I(\bkappa;\nu), \label{ModIndex}
\end{eqnarray}
where ${\bf v}$ is the velocity of the interstellar medium transverse to the line of sight to the source.
A typical value of $v=30\,$km\,s$^{-1}$ is derived from direct measurements using intra-day variable radio sources (Dennett-Thorpe \& de Bruyn 2003; Bignall et al. 2003).

The foregoing theory applies to the autocorrelation of the intensity fluctuations at a single frequency.  However, as we are interested in comparing the intensities of spectral features relative to one another and to the continuum, we also compute the covariance between intensity fluctuations at two distinct frequencies.

The frequency dependence of the covariance is potentially influenced by both its dependence on the scintillation power spectrum, $P_{\rm scint}(\bkappa)$, and on the variation in image structure with frequency.   
However, the scintillations themselves are broadband (e.g. Goodman \& Narayan 1989), which is to say that a points source exhibits similar intensity fluctuations over a large range of observing frequencies $\Delta \nu \sim \nu$.  Variations in $P_{\rm scint}(\bkappa)$ with frequency are therefore negligible for present purposes, and it suffices to consider only the influence of variation in image structure with frequency.
The cross power between observations at frequencies $\nu$ and $\nu'$ is (Little \& Hewish 1966; Rickett 2000)
\begin{eqnarray}
P_I(\bkappa;\nu;\nu') = P_{\rm scint}(\bkappa) V_{\rm app}\left( \frac{\bkappa L}{k} ;\nu \right)  \,  V_{\rm app}^*\left( \frac{\bkappa L}{k} ;\nu' \right) \label{CrossPower} 
\end{eqnarray}
This is the generalisation of Eq.\,(\ref{PCentral}), with the power spectrum of the image at frequency $\nu$ replaced by the cross power spectrum of the images at $\nu$ and $\nu'$.  The corresponding temporal cross-correlation of the intensity fluctuations is
\begin{eqnarray}
C_I(t;\nu,\nu') = \langle \Delta I(t';\nu) \Delta I(t'+t;\nu') \rangle =  \int d^2 \bkappa \, e^{i \bkappasm \cdot {\bf v}t} P_I(\bkappa;\nu,\nu'). \label{Covariance}
\end{eqnarray}
The auto- and cross-correlations given in Eqs. (\ref{ModIndex}) and (\ref{Covariance}) relate the scintillation properties to the image structure.

\begin{figure*}
\centering
\includegraphics[width=15cm]{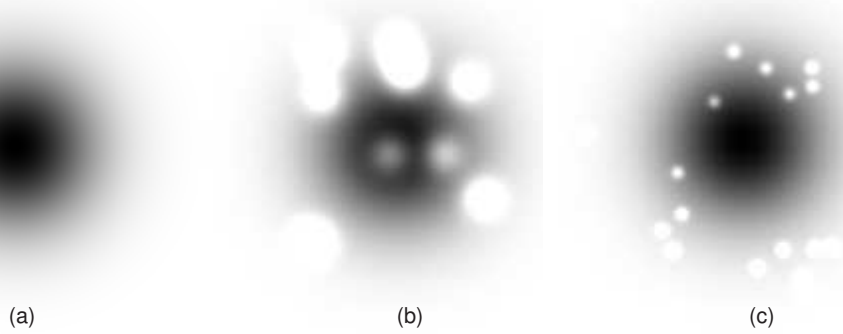} 
\caption{The variability due to the continuum image of a source (a) is different relative to the image of the source at two spectral lines at frequencies, say, (b) $\nu_{\rm abs1}$ and (c) $\nu_{\rm abs2}$.   Since the observed variability properties depend sensitively on fine-scale source structure, the intensity scintillations caused by scattering in the ISM of our Galaxy are different in all three cases.
}
\label{HIimages}
\end{figure*}

\subsection{Assessment of variability relative to the continuum}

Spectral line variations can occur in sources whose continuum level also varies.  Since only variations relative to the continuum indicate the presence of substructure in the absorbing material it is necessary to construct a measure of the variability that removes the contribution from continuum variability.

It is possible to separate spectral line variations from continuum fluctuations by computing how the flux density of an absorption line would vary due solely to changes in the continuum level.  
The fluctuations of a spectral line whose flux density varies proportional to the continuum level, $\Delta I_{\rm dep}(\nu_{\rm abs})$ are 
\begin{eqnarray}
\Delta I_{\rm dep}(\nu_{\rm abs}) = e^{-\tau} \Delta I (\nu) = \left( \frac{S_{\rm src}-S_{\rm abs}}{S_{\rm src}} \right) \, \Delta I (\nu),
\label{FluctScaling}
\end{eqnarray}
where $\tau$ is the optical depth, $\Delta I(\nu)$ are the continuum flux density variations, $S_{\rm src}$ is the mean source continuum flux density, and $S_{\rm abs}$ is the mean absorbed flux density at the frequency $\nu_{\rm abs}$ of the absorption line.  

Variations induced by scintillation that are independent of the continuum level might be mistaken for variations in the optical depth, $\tau$.
The amplitude of spectral line fluctuations independent of the continuum fluctuations is given by $\Delta I_{\rm indep} (\nu_{\rm abs}) \equiv \Delta I(\nu_{\rm abs}) - \Delta I_{\rm dep}(\nu_{\rm abs})$.  
The auto-correlation of independent intensity fluctuations is then 
\begin{eqnarray}
C_{\rm indep}(\nu_{\rm abs}) &\equiv& \left\langle \Delta I_{\rm indep}(t', \nu_{\rm abs}) \Delta I_{\rm indep}(t'+t,\nu_{\rm abs}) \right\rangle \nonumber \\
&=& C_I(t;\nu_{\rm abs},\nu_{\rm abs}) - \left(\frac{S_{\rm src}-S_{\rm abs}}{S_{\rm src}} \right) \left[ C_I(t;\nu_{\rm abs},\nu) + C_I(t;\nu,\nu_{\rm abs})\right]+   \left(\frac{S_{\rm src}-S_{\rm abs}}{S_{\rm src}} \right)^2 C_I(t;\nu,\nu).  \label{IndepIVar}
\end{eqnarray}
This expression subtracts out the continuum variability contribution to the variations observed in the spectral line.  It is zero for any source whose spectral features vary identically to the continuum fluctuations, as would be the case if the image of the source at $\nu_{\rm abs}$ were identical to its continuum image, if the absorption structure were distributed homogeneously. 

A similar quantity assesses the covariance of independent variations between two spectral lines.  For lines at frequencies $\nu_{\rm abs1}$ and $\nu_{\rm abs2}$ with absorbed flux densities $S_{\rm abs1}$ and $S_{\rm abs2}$ the cross-correlation is
\begin{eqnarray}
C_{\rm indep}(t;\nu_{\rm abs1},\nu_{\rm abs2}) &\equiv& \left\langle \Delta I_{\rm indep}(t', \nu_{\rm abs1}) \Delta I_{\rm indep2}(t'+t,\nu_{\rm abs}) \right\rangle \nonumber \\
&=& C_I(t,\nu_{\rm abs1},\nu_{\rm abs2}) 
+ \left(\frac{S_{\rm src}-S_{\rm abs1}}{S_{\rm src}} \right) 
\left(\frac{S_{\rm src}-S_{\rm abs2}}{S_{\rm src}} \right) C_I(t,\nu,\nu) \nonumber \\ &\null&
- \left(\frac{S_{\rm src}-S_{\rm abs2}}{S_{\rm src}} \right)  C_I(t,\nu_{\rm abs1},\nu) 
- \left(\frac{S_{\rm src}-S_{\rm abs1}}{S_{\rm src}} \right) C_I(t,\nu,\nu_{\rm abs2}) 
.  \label{IndepICovar}
\end{eqnarray}
This quantity removes fluctuations in the source continuum flux density when the covariance in the intensity fluctuations between two spectral features is computed.  Again, it is identically zero if a spectral line and the continuum exhibit identical fractional flux density variations.

\subsubsection{Relevant angular scales}
Before applying the foregoing theory to specific absorption distributions, it is appropriate to consider the range of scales probed by scintillation.  Scintillation is most sensitive to image structure on angular scales comparable to $\theta_{\rm ref} = r_{\rm ref}/L$ (viz. Eq.\,(\ref{Codona11})) .  The refractive angular scale is expressed in terms of standard scintillation parameters as follows
\begin{eqnarray}
\theta_{\rm ref} = 2.15 \,\left( \frac{C_N^2}{10^{-4}\,{\rm m}^{-20/3} } \right)^{3/5} \left( \frac{L}{0.5\,{\rm kpc}} \right)^{3/5} \left( \frac{\lambda}{1\,{\rm m}} \right)^{11/5}\, {\rm mas},
\label{ThetaRef} 
\end{eqnarray}
where we have equated the thickness of the scattering material with the distance to the scattering screen, $\Delta L = L$.
For instance, scintillation of a spectral line due to neutral hydrogen at $z=0.1$ (at $\lambda\,0.232\,$m) probes structure on a spatial scale of $\sim 2\,$pc, for typical scattering parameters $C_N^2 = 10^{-4}\,$m$^{-20/3}$ and $L=0.5\,$kpc for a line of sight off the Galactic plane.  The response of scintillation to structure on scales larger than $\theta_{\rm ref}$ decreases sharply.  The modulation index of a source of size $\theta$ scales as 
$(\theta/\theta_{\rm ref})^{\beta/2 -3}$ once its size exceeds $\theta_{\rm ref}$.

\section{Absorption Models}\label{applications}
We now apply the foregoing results to two hypothetical distributions of absorbing material:  one a population of clouds with fixed angular size, the other a medium whose power spectrum of absorption brightness fluctuations follows a power law.   Absorption against both extended and compact sources is considered.

The models are also applicable to variability from systems in emission.  A distribution of material in emission would exhibit the same variability properties as the {\it same} distribution seen in absorption against an extended background source.  This is because the power spectra of the angular brightness distributions in the two cases are identical, except for the contribution at zero spatial frequency ($\bkappa=0$) due to the background source.  As mentioned above, this contribution does not influence the variability.

\subsection{Structure with a Characteristic Angular Scale}

\subsubsection{Emission, or absorption against an extended background source}\label{extendedGauss}
Consider an extended background source whose spectrum is altered by a single intervening system.  The intensity of the background source, $I_0$, is assumed constant over an angular scale much greater than the refractive scale $\theta_{\rm ref}$, so that it does not scintillate.

We consider a model in which the intervening system is comprised of two sets of clouds, with $N_1$ clouds of size $\theta_1$ absorbing at centre frequency $\nu_1$ and $N_2$ of size $\theta_2$ absorbing at centre frequency $\nu_2$.   This is the simplest model which demonstrates that different spectral lines can vary relative to one another.  The generalisation to an arbitrary number of spectral lines and cloud types is obvious, but introduces no new physics.

To aid in the computation of scintillation effects the exponential of the optical depth is specified in the following {\it idealised} form
\begin{eqnarray}
\exp[-\tau(\btheta;\nu)]  &=&  \,\,\,\,\,\,
 \left[ 1-  A_1 \varphi(\nu-\nu_1) \sum_i^{N_1} \exp\left( -\frac{(\btheta-\Delta \btheta_{1,i})^2}{\theta_1^2}\right) \right] 
+ 
 \left[ 1-  A_2 \varphi(\nu-\nu_2) \sum_j^{N_2} \exp\left( -\frac{(\btheta-\Delta\btheta_{2,j})^2}{\theta_2^2}\right) \right], \nonumber \\
\end{eqnarray}
where the angular offsets $\Delta \btheta_{1,i}$ and $\Delta \btheta_{2,i}$ describe the positions of the cloud centres on the sky.  The absorption profile is described by the function $\varphi(\nu)$, normalised such that the absorbed flux of a source of unit flux density is unity, $\int d\nu \, \varphi(\nu) =1$.  We make the simplifying assumption that the line profiles at frequencies $\nu_1$ and $\nu_2$ do not overlap.  This ensures that the scintillations observed at frequency $\nu_1$ are entirely due to one set of absorbing clouds, while the scintillations observed at $\nu_2$ are due to another set.  The absorption strengths $A_{1,2}$ determine the amplitude of the line, with $0< A_{1,2} < 1$.  The image visibility is then
\begin{eqnarray}
V_{\rm app}({\bf r};\nu) &=&  \frac{4 \pi^2 I_0 \delta^2({\bf r})}{k^2} -  I_0 \sum_{\eta=1,2} A_\eta \pi \theta_\eta^2 \exp \left( - \frac{k^2 \theta_\eta^2 r^2 }{4} \right) \varphi(\nu-\nu_\eta) \sum_i^{N_\eta} \exp \left[ - i k {\bf r} \cdot \Delta \btheta_{\eta,i} \right],
\end{eqnarray}
where the two cloud types are indexed by $\eta$. 
The first term represents the contribution from the background source, assumed to have a spatially constant flux density.  This term only contributes at ${\bf r}=0$, and is henceforth neglected since it has no influence on the power spectrum of intensity fluctuations (see Eq.\, (\ref{Codona11})).


When the number of absorbing clouds is large one is justified in averaging over their positions.
The power spectrum of the apparent image contains an average over the cloud positions $\langle \exp\left[ -i k {\bf r} \cdot (\Delta \btheta_{\eta,i} - \Delta \btheta_{\eta,j})\right] \rangle$ with $\eta=[1,2]$ and $i,j=[1,\ldots,N_{\eta}]$.  The evaluation of this quantity has been treated in a similar context elsewhere (Macquart 2004), to which we refer the reader for more detail.

It is convenient to separate this average into the $N_\eta$ self-terms and $N_\eta(N_\eta-1)$ cross-terms.
Each self-term yields a contribution of unity.  The average over the cross terms proceeds by considering the clouds to be distributed randomly over an angular area $\Omega$, with $p_{2} (\btheta_i,\btheta_j)/\Omega^2$ being the joint probability of finding the $i$th cloud at position $\btheta_i$ and the $j$th cloud at position $\btheta_j$.  We assume that the statistics of the cloud positions {\bf vary} only weakly over the area in which the clouds are distributed, so that we may write the probability distribution as a function of object separation, $\Delta \btheta=\btheta_i-\btheta_j$ only, and neglect its dependence on the average cloud location, ${\bf \Theta}=(\btheta_i+\btheta_j)/2$.  Thus one has
\begin{eqnarray}
\sum_{j,l}^{N_\eta} \left\langle \exp\left[ -i k {\bf r} \cdot (\Delta \btheta_{\eta,j} - \Delta \btheta_{\eta,l})\right] \right\rangle
&=& N_\eta +  \frac{N_\eta (N_\eta -1)}{\Omega^2} \int d^2 {\bf \Theta} d^2 \Delta \btheta \, p_2( \Delta \btheta) \exp[i k {\bf r} \cdot \Delta \btheta] \nonumber \\ 
&=& N_\eta + \frac{N_\eta (N_\eta -1)}{\Omega} \frac{\tilde p_2(k{\bf  r})}{\Omega},
\end{eqnarray}
where $\tilde p_2$ is the Fourier transform of $p_2$.

The average power spectrum of the brightness distribution is 
\begin{eqnarray}
\left\langle \left\vert V_{\rm app}({\bf r};\nu)\right\vert^2 \right\rangle &=& \sum_{\eta=1,2}  \frac{S_{\rm abs \eta}^2}{N_\eta}  \left[ 1 + \frac{N_\eta-1}{\Omega} \tilde p_2(k {\bf r}) \right] \exp \left( - \frac{k^2 \theta_\eta^2 r^2 }{2} \right) \varphi^2(\nu-\nu_\eta),  \label{AvgPower}
\end{eqnarray}
where ${S_{\rm abs}}_{\eta}=A_{\eta} \pi \theta_{\eta}^2 N_{\eta} I_0$ are the flux densities absorbed by each population of clouds.  

In the simple case in which the clouds are sparsely distributed (i.e. $N_\eta/\Omega<1$), it is convenient to neglect the effect of the term involving $\tilde p_2$.  In fact, this term may always be ignored when the region over which the clouds are distributed greatly exceeds the refractive angular scale.  In this case $\tilde p_2({\bf r}) $ falls to zero at sufficiently small $r$ that it yields no contribution to the scintillation variations (cf. Eq. (\ref{ModIndex})).  This is because the term involving $\tilde p_2$ only contains power on angular scales much larger than that probed by scintillation.


We now evaluate the auto-correlation of the temporal intensity fluctuations due to scintillation.  We make the simplifying assumption that the angular scale of the absorbers is greater than or comparable to the refractive scale $\theta_{1,2} > \theta_{\rm ref}$, which enables us to write the auto-correlation function in closed form (Coles et al. 1987).  In this case the source power spectrum cuts off the integral in Eq. (\ref{ModIndex}) over $\bkappa$ before the exponential term involving $D_\phi$ (see Eq. (\ref{Codona11})) becomes important and while the argument of the sine function is still small.  Evaluation of the integral yields
\begin{eqnarray}
C_{\rm indep}(t;\nu) &=& 4 \pi^2 r_e^2 \lambda^2 {\rm SM} \left( \frac{L}{k} \right)^2 2^{2-\beta/2} \Gamma \left(3-\frac{\beta}{2} \right) 
\sum_{\eta=1,2}  \varphi^2(\nu-\nu_\eta) \frac{ {S_{\rm abs}}_\eta^2 }{N_\eta}  
(L \theta_\eta)^{\beta-6} \nonumber \\
&\null& \qquad \qquad \qquad \qquad \qquad \qquad \qquad \qquad \qquad \qquad \times \null_1 F_1 \left[3-\frac{\beta}{2},1,-\frac{v^2 t^2}{2 L^2 \theta_\eta^2}\right] . \label{CILargeSrc}
\end{eqnarray}
The variance of the intensity fluctuations is obtained by setting $t=0$.  

In this case a numerical estimate for the root mean square flux density variation when $\beta=11/3$ is 
\begin{eqnarray}
\langle (\Delta I)^2 \rangle_{\eta}^{1/2} &=& 0.36 \, \varphi(\nu-\nu_\eta) 
\, \left(\frac{ C_N^2}{10^{-4} \,{\rm m}^{-20/3} }\right)^{1/2} 
\left( \frac{L}{0.5\,{\rm kpc}} \right)^{1/3} \left( \frac{\lambda}{1\,{\rm m}} \right)^{2}  \left( \frac{\theta_\eta}{1\,{\rm mas} } \right)^{-7/6}  
\left[ \frac{{S_{\rm abs}}_{\eta}^2}{N_{\eta}} \right]^{1/2}. \label{RMSI}
\end{eqnarray}
For a fixed absorbing flux density, the modulation index decreases as 
$N_\eta^{-1/2}$.  This is because the observed intensity fluctuations may be regarded, in a statistical sense, as the scintillation pattern from $N_\eta$ independent absorbing clouds.  The assumption that the absorbers are distributed homogeneously in a statistical sense over the area subtended by the background source ensures that the scintillation patterns due to individual absorbers are indepedent  when averaged over the absorbing population.
For a fixed number of absorbing clouds, Eq.\,(\ref{RMSI}) implies that large absorbing clouds contribute the most to intensity variations.  The root mean square intensity fluctuation is proportional to 
$N_\eta^{-1/2} \theta^{-7/6} S_{\rm abs} \propto N_\eta^{1/2} \theta^{5/6}$.  Thus the number of clouds of size $\theta$ must rise faster than $N_\eta \propto \theta^{-5/3}$ for decreasing $\theta$ for the contributions of small clouds to be equally important as those of the largest clouds.  Otherwise, only the largest clouds in the absorbing medium contribute substantially to the observed spectral variability.

We also derive the time scale of fluctuations, defined as the time scale at which the correlation function reaches $1/e$ of its maximum value, from Eq.\,(\ref{CILargeSrc})
\begin{eqnarray}
{t_{\rm scint}}_\eta = 36.9 \left( \frac{\theta_\eta}{1\,{\rm mas}} \right) \left( \frac{L}{0.5\,{\rm kpc}} \right) \left( \frac{v }{30\,{\rm km\,s}^{-1} } \right)^{-1} {\rm days}.
 \end{eqnarray}
This time scale depends only on the size of the absorbers assuming $\theta_\eta > \theta_{\rm ref}$. The ratio of fluctuation time scales of any two spectral lines is equal to the relative sizes of the absorbing objects, $t_{\nu_1}/t_{\nu_2} = \theta_1/\theta_2$.

The cross-correlation of the intensity fluctuations $C_{\rm indep}(t,\nu_1,\nu_2)$ is determined by the cross power of the visibilities measured at $\nu_1$ and $\nu_2$.   This depends on the distribution of the positions of clouds of the two different sizes, $\theta_1$ and $\theta_2$.  A full description of the intensity cross-correlation involves the joint probability distribution of cloud centres between the two cloud types.  However, a practical simplifying assumption is that the two  types of clouds are distributed independently.  If we assume that the cloud positions follow identical distribution functions, with the probability of finding a cloud $i$ at position $\btheta_i$ being $p_1(\btheta_i)/\Omega$, then the cross-correlation of the intensity fluctuations is proportional to 
\begin{eqnarray}
\left\langle \sum_{i,j} \exp \left[i  k {\bf r} \cdot (\btheta_{1,i} - \btheta_{2,j}) \right] \right\rangle =  \frac{N_1 N_2}{\Omega^2} 
| \tilde p_1(k {\bf r}) |^2. \label{pCross}
\end{eqnarray}
Now if the clouds are distributed out to some maximum scale $\theta_{\rm max}$ in the absorbing region, the quantity $| \tilde p_1(k {\bf r}) |^2$ declines rapidly for $r > \theta_{\rm max}^{-1}$.
In any realistic medium the absorbing clouds are distributed over an angular extent much larger than the angular scale probed by refractive scintillation (i.e. $\theta_{\rm max} \gg \theta_{\rm ref}$).  Thus the visibility cross-power, $V_{\rm app}(\bkappa L/k,\nu_1) V_{\rm app}^*(\bkappa L/k, \nu_2)$, is cut off at wavenumbers $\kappa > (L \theta_{\rm max})^{-1}$.  This wavenumber is much smaller than those probed by interstellar scintillation, $\kappa \sim r_{\rm ref}^{-1}$ (viz. Eq. (4)).  This implies that the covariance of the intensity fluctuations, as described in Eqs. (\ref{Covariance}) and (\ref{IndepICovar}), is zero.

In physical terms, the result in Eq. (\ref{pCross}) shows that the cross-power of the visibilities at the two spectral lines exhibits structure only on angular scales comparable to the entire region over which the clouds are distributed.  As this scale is much larger than that probed by interstellar scintillation, it does not respond to the image structure.   This is the reason that the cross-correlation between spectral lines emanating from the two independently distributed sets of absorbing clouds is zero.

In summary, we have shown that spectral lines due to absorbing clouds randomly distributed in front of a large background source can exhibit substantial variations due to interstellar scintillation. 
However, when multiple spectral lines are present the variations between the lines are not correlated under most conditions.




\subsubsection{Absorption against a compact background source} 
If the background source is itself compact enough to scintillate this has two important implications for the spectral variability: (i) the continuum level also varies, so that the assessment of spectral variability needs to be made relative to the varying continuum, and (ii) the radiation probes a only narrow cone of HI structure along the line of sight to the observer, so that only the absorption of relatively few clouds need be considered.  
%
%
To illustrate the effect of background source size on spectral variability we consider a background source with a Gaussian intensity distribution
\begin{eqnarray}
I_{\rm src}(\btheta)=I_0 \exp \left[ - \frac{\btheta^2}{\theta_{\rm src}^2}\right],
\label{Isrc}
\end{eqnarray}
whose angular size is sufficiently small that it exhibits scintillation.  The source background flux density is $S_{\rm src}=I_0 \pi \theta_{\rm src}^2$.  We consider absorption from the same intervening system 
of absorbing clouds as above, so the power spectrum and cross powers of the absorbed image are
\begin{eqnarray}
V_{\rm app}({\bf r};\nu) V_{\rm app}^*({\bf r};\nu) &=& S_{\rm src}^2 \left\{ \exp \left[ -\frac{k^2 r^2 \theta_{\rm src}^2}{4} \right] - G({\bf r}) \right\}^2, \label{VappPower} \\
V_{\rm app}({\bf r};\nu) V_{\rm app}^*({\bf r};\nu') &=&  \, S_{\rm src}^2 
\exp \left[ - \frac{ k^2 r^2 \theta_{\rm src}^2}{4}\right] \left\{ 
 \exp \left[ - \frac{k^2 r^2 \theta_{\rm src}^2}{4}\right]  - G({\bf r}) \right\}, \label{VappCross}
\end{eqnarray}
where,
\begin{eqnarray}
G({\bf r}) = \sum_{\eta=1,2} A_\eta  \frac{ \theta_\eta^2}{\theta_{\rm src}^2 + \theta_\eta^2} \varphi(\nu-\nu_\eta) \sum_j^{N_\eta} 
\exp \left[ - \frac{k^2 r^2 \theta_{\rm src}^2 \theta_\eta^2 + 4 i k \theta_{\rm src}^2 {\bf r} \cdot \Delta \btheta_{\eta,j} + 4 \Delta \btheta_{\eta,j}^2}{4(\theta_{\rm src}^2 + \theta_\eta^2)} \right].
\end{eqnarray}



Variability from a collection of absorbers is treated in Appendix \ref{ManyAbsorbers}, but in many practical cases it suffices to consider the intensity modulation of a background source with only a single absorber directly in front of it ($\Delta \btheta =0$).  If both the absorber and background source are large, $\theta_{\rm src}, \theta_{\rm 1} \ga \theta_{\rm ref}$, we can use the same approximation made to derive Eq. (\ref{CILargeSrc}).  The intensity auto- and cross-correlations are
\begin{subequations} 
\begin{eqnarray}
C_I({\bf r};\nu) &=& 4 \pi^2 r_e^2 \lambda^2 \,{\rm SM} \left(\frac{L}{k} \right)^2 2^{2-\beta/2} \Gamma \left( 3-\frac{\beta}{2} \right) S_{\rm src}^2 \left\{ (L \theta_{\rm src})^{\beta-6} \null_1 F_1 
\left( 3-\frac{\beta}{2},1,-\frac{r^2}{2L^2 \theta_{\rm src}^2 }\right) \right. \nonumber \\
&\null& \left. \qquad \qquad 
- 2  A_1 \,\varphi(\nu-\nu_1)\, 
(L {\theta_{\rm cross}}_1)^{\beta-6} \frac{\theta_1^2}{\theta_{\rm src}^2 + \theta_1^2} \null_1 F_1 \left(3 -\frac{\beta}{2},1,-\frac{r^2}{2L^2 {\theta_{\rm cross}}^2} \right) \right. \nonumber \\
&\null& \left. \qquad \qquad + 
 A_1^2 \, \varphi(\nu-\nu_1) \, 
(L \theta_{\rm eff})^{\beta-6} \left( \frac{\theta_1^2}{\theta_1^2 + \theta_{\rm src}^2} \right)^2 \null_1 F_1 \left(3-\frac{\beta}{2},1,- \frac{r^2}{2 L^2 {\theta_{\rm eff}}^2} \right)
\right\} , \label{SmallSrcModIndex}  \\
\hbox{and} \nonumber \\
C_I({\bf r};\nu,\nu') &=&  4 \pi^2 r_e^2 \lambda^2 {\rm SM} (L/k)^2 2^{2-\beta/2} \Gamma \left(3- \frac{\beta}{2} \right)  S_{\rm src}^2 \, \varphi(\nu'-\nu_1)\, \left\{   
(L \theta_{\rm src})^{\beta-6} \null_1 F_1 \left(3-\frac{\beta}{2},1,-\frac{r^2}{2 L^2 \theta_{\rm src}^2}\right) \right.  \nonumber \\
&\null& \left. \qquad \qquad \qquad 
-  A    \, \left(L \theta_{\rm cross} \right)^{\beta-6} \,\frac{\theta_1^2}{\theta_{\rm src}^2 + \theta_1^2} \,
\null_1 F_1 \left( 3-\frac{\beta}{2},1,- \frac{r^2}{2 L^2 \theta_{\rm cross}^2}\right) \right\}, \label{CrossILargeSrc}
\end{eqnarray}
\end{subequations}
respectively where, 
\begin{subequations}
\begin{eqnarray}
\theta_{\rm cross}^2 &=& \frac{\theta_{\rm src}^2(2 \theta_1^2 + \theta_{\rm src}^2)}
{2(\theta_1^2+\theta_{\rm src}^2)}, \\
\theta_{\rm eff}^2 &=& \frac{\theta_1^2 \theta_{\rm src}^2}{
\theta_1^2 + \theta_{\rm src}^2}
\end{eqnarray}
\end{subequations}
are effective angular diameters.  The quantity $\theta_{\rm cross}^2$ tends to $\theta_{\rm src}^2/2 $ when the background source angular diameter exceeds that of the absorber in the limit $\theta_{\rm src} \gg \theta_\eta$, and tends to $\theta_{\rm src}^2$ in the opposite limit.  Equation (\ref{SmallSrcModIndex}) provides a means to compute the modulation index of the continuum fluctuations, line fluctuations and the ``independent'' line fluctuations.

The time scale of the intensity cross correlation, set by $\theta_{\rm src}$ and $\theta_{\rm cross}$, is insensitive to the angular scales of the absorbers if they are smaller than the background source, since $\theta_{\rm cross} \approx \theta_{\rm src}/\sqrt{2}$ is insensitive to $\theta_1$ in this regime.

\subsubsection{Clumpy cloud distribution} \label{Clumpy}
Equations (\ref{VappPower}) and (\ref{VappCross}) simplify when the positions of the clouds $\Delta \btheta_{\eta,j}$ do not greatly exceed the background source size.  In this case, the terms $\Delta \btheta_{\eta,j}^2$ in the exponentials of these two equations are negligible, and an average over random clouds centres yields
\begin{eqnarray}
\langle V_{\rm app}({\bf r};\nu) V_{\rm app}^*({\bf r};\nu) \rangle &=& S_{\rm src}^2 \left\{ \exp \left[ -\frac{k^2 r^2 \theta_{\rm src}^2}{2} \right] 
+ \sum_{\eta=1,2}  \varphi(\nu-\nu_\eta) A_\eta^2 \left(\frac{ \theta_\eta^2}{\theta_{\rm src}^2 + \theta_\eta^2}\right)^2  N_\eta 
\exp \left[ - \frac{k^2 r^2 \theta_{\rm src}^2 \theta_\eta^2}{2(\theta_{\rm src}^2 + \theta_\eta^2)} \right] \right\},
\label{VappPowerSimple}
\end{eqnarray}
and 
\begin{eqnarray}
\langle V_{\rm app}({\bf r};\nu) V_{\rm app}^*({\bf r};\nu') \rangle &=&  \, S_{\rm src}^2 
\exp \left[ - \frac{k^2 r^2 \theta_{\rm src}^2}{2}\right]. \label{VappCrossSimple}
\end{eqnarray}
The intensity auto- and cross-correlations are then respectively
\begin{subequations}
\begin{eqnarray}
C_I(t;\nu) &=& m_{\rm ref}^2 S_{\rm src}^2 \left\{ 
\left( \frac{\theta_{\rm src}}{\theta_{\rm ref}} \right)^{\beta-6} 
	\null_1 F_1 \left( 3-\frac{\beta}{2},1,-\frac{v^2 t^2}{2L^2 \theta_{\rm src}^2 }\right) \right. \nonumber \\
&\null&  \left. + \sum_{\eta=1,2} \varphi (\nu-\nu_\eta) A_\eta^2 N_\eta 
\left( \frac{\theta_{\rm eff \eta}}{\theta_{\rm ref}}\right)^{\beta-6} \left( \frac{\theta_\eta^2}{\theta_\eta^2 + \theta_{\rm src}^2} \right)^2 \null_1 F_1 \left(3-\frac{\beta}{2},1,- \frac{v^2 t^2}{2 L^2 {\theta_{\rm eff \eta}}^2} \right)
\right\} \label{CInu}
\end{eqnarray}
and
\begin{eqnarray}
C_I(t;\nu,\nu') &=& m_{\rm ref}^2 S_{\rm src}^2 \, \null_1 F_1 \left( 3-\frac{\beta}{2},1,-\frac{v^2 t^2}{2L^2 \theta_{\rm src}^2 }\right),
\end{eqnarray}
where the quantity
\begin{eqnarray}
m_{\rm ref}^2 = 2^{1-\beta/2} (\beta-2) \left( 2-\frac{\beta}{2} \right) \Gamma \left( \frac{\beta}{2} \right) \left( \frac{r_{\rm diff}}{r_{\rm F}} \right)^{8-2\beta} 
\end{eqnarray}
\end{subequations}
approximates the variance that a point-like source of unit intensity would exhibit.  The amplitude of the intensity variations exhibited by a spectral line, given by setting $t=0$ in Eq. (\ref{CInu}), contains a component that is due to the size of the background source, and another component which scales linearly with the number of absorbing objects, $N_\eta$.  Note, however, that the amplitude of the cross-correlation between spectral frequencies $\nu$ and $\nu'$ is independent of the characteristics of the absorbers, and is merely equal to that which a continuum source would exhibit.

\subsection{A statistical description of emission or absorption fluctuations}
Instead of regarding the absorbing (or emitting) medium to be composed of discrete clouds, as discussed above, it is often more useful to consider the optical depth as a stochastic variable which is allowed to vary over the angular extent of the background source.  In this section we derive the properties of spectral line variability caused by an absorbing medium whose spatial variations are specified only in a statistical sense.  Spectral line variability in this model is caused only by spatial optical depth variations, which in turn may induce structure in the image of the absorbed source on sufficiently small scales to be subject to interstellar scintillation.  The absorbing medium is not regarded as moving across the extent of the source itself, so no spectral variability is caused by relative motion of the absorbing material across the line of sight to the source.  We assume that the optical depth variations associated with a given spectral line are confined to a thin screen a distance $D_{\rm abs}$ from the scattering medium in which the scintillations take place.  

In general terms one can describe the effect of the local medium on the incident electric field of the background source, $u_{\rm src}({\bf r})$, as follows
\begin{eqnarray}
u({\bf r}) = u_{\rm src}({\bf r}) e^{i \phi ({\bf r})}, \label{uTauReln} 
\end{eqnarray}
where the phase factor $\phi$ is regarded as a complex number.  The quantity Re($\phi$) is the phase delay caused by refraction, and Im($\phi)=\tau/2$ represents the effect of absorption (or amplification) by the medium\footnote{Note that Eq. (\ref{uTauReln}) reduces to the familiar result $I({\bf r}) =u({\bf r}) u^*({\bf r}) = I_{\rm src}({\bf r}) \,e^{-\tau ({\bf r})}$.}.  For present purposes we are only interested in optical depth variations, so we henceforth set Re($\phi)=0$ and consider only fluctuations in the quantity $\eta \equiv$ Im($\phi)$.  

Two specific assumptions are made concerning the statistical nature of the optical depth fluctuations.  Firstly, it is assumed that the fluctuations are wide sense stationary, so that the statistical properties of $\tau({\bf r})$ do not depend on the position.  Specifically, this assumption implies that quantities involving changes in optical depth between positions say, ${\bf r}'+{\bf r}$ and ${\bf r}'$, depend only on their separation, ${\bf r}$.  

The second assumption is that the statistical fluctuations in the optical depth follow a f distribution.  We make this assumption in lieu of more detailed observational measurements of the distribution of optical depth fluctuations for any given spectral line.  The assumption of Gaussian statistics has proven highly successful in treating fluctuations in the electron density variations in the ionized component in the interstellar medium of our Galaxy, which give rise to variations in Re$(\phi$) (see, for example, Armstrong, Rickett \& Spangler 1995).  

However, the model contains an implicit shortcoming, because normally distributed random variates are not constrained to either positive or negative values.  As such, it makes no distinction between the statistical properties of the spatial distribution of $\tau$ depending on whether it is positive or negative.
Nonetheless, this shortcoming is not serious if the optical depth variations are small compared to their mean value, so that most of the medium is seen chiefly in either absorption or emission.

Recall from the discussion in Sect. 2.2 that the scintillation characteristics of the absorption (or emission) lines depend on the mean square visibility of the absorbed (or emitting) image.  We derive a general result for this quantity in Appendix \ref{TauDeriv}, and merely quote the result here:
\begin{eqnarray}
\langle |V_{\rm app} ({\bf r})|^2 \rangle = |V_{\rm src}({\bf r}) |^2 \, \exp\left[ -2 \langle \tau \rangle + \frac{3}{2} C_\tau(0) - D_\tau({\bf r}) \right], \label{VGeomOpt}
\end{eqnarray}
where $V_{\rm src}({\bf r})$ is the visibility of the background source, $C_\tau({\bf r})=\langle [\tau({\bf r}+{\bf r}')-\langle \tau \rangle][\tau({\bf r})-\langle \tau \rangle] \rangle$ is the autocovariance of the optical depth fluctuations in the absorbing medium and $D_\tau({\bf r}) =2 [C_\tau(0) - C_\tau({\bf r})]$ is its corresponding structure function.  
Equation (\ref{VGeomOpt}) is correct in the limit of geometric optics, when the opacity variations occur on a scale that is large compared to $D_{\rm abs}/k$, which is likely to be the case in practice.  However, Appendix \ref{TauDeriv} also contains a more general expression for the mean square visibility when wave optics are important.

The physical significance of Eq. (\ref{VGeomOpt}) is that the opacity variations increase the amount of fine-scale structure in the image of the absorbed source.  This is evident through the effect of the quantity $\exp\{[3 C_\tau(0)- D_\tau ({\bf r})]/2\}$.  The term involving $C_\tau(0)$ boosts the mean square visibility of the image by an amount that is independent of baseline length, ${\bf r}$.  However, the influence of the $D_\tau ({\bf r})$ partially offsets this effect, so that there is an overall enhancement in the mean square visibility by a factor of $\exp[C_\tau (0)/2]$ for very large baselines.

The structure function of opacity variations is directly related to the power spectrum of  optical depth variations, $P_\tau(\kappa)$, by a Fourier transform:
\begin{eqnarray}
D_\tau({\bf r}) = 2 \int d^2 \bkappa \, [1- \exp(i \bkappa \cdot {\bf r})] P_\tau(\bkappa). 
\end{eqnarray}

The power spectrum of optical depth variations is well known for some spectral lines.  For instance, recent studies of the Milky Way, Large and Small Magellanic clouds, and nearby galaxies (Braun 1999, Lazarian \& Pogosyan 2000) show that the power spectrum of HI emission fluctuations\footnote{As we refer to fluctuations in the plane of the sky, the argument of the power spectrum is a two-dimensional vector.  This power spectrum depends on the manner in which the averaging over radial distance (or radial velocity) has been performed.  The inversion from the full three-dimensional HI density field to the power spectrum of optical depth variations is discussed at length in Lazarian \& Pogosyan (2000).} follows a power law.  Green (1993) finds the index of the power law to be in the range $n=2-3$, Deshpande, Dwarakanath \& Goss (2000) find $n \sim 2.75$.  Lazarian \& Pogosyan (2000) show that the index of the power law found depends on the spectral resolution of the observations, and on whether any averaging has been performed over spectral channels.  These authors tentatively conclude that the density fluctuations follow a Kolmogorov ($n=11/3$) power law.  Observations of HI absorption also suggest that the structure follows a power law, even out to scales comparable to the size of the absorbing galaxy (e.g. Braun 1999).  Motivated by this observational evidence, we consider optical depth variations from a power spectrum of the form 
\begin{eqnarray}
P_\tau (\bkappa) = Q_0 | \bkappa|^{-n} \exp \left[ -\left( \frac{\bkappa}{2 \bkappa_{\rm max}} \right)^2 \right],
\end{eqnarray}
where $q_{\rm max}=l_{\rm min}^{-1}$ is the high-wavenumber cutoff.  In a turbulent medium, this cutoff physically corresponds to an inner dissipation scale, $l_{\rm min}$, associated with the turbulent cascade. The structure function for $n<4$ and $r \gg l_{\rm min}$ takes the form
\begin{eqnarray}
D_\tau ({\bf r}) &=& 
\left( \frac{r}{r_\tau} \right)^{n-2}, \qquad \hbox{with } r_\tau = \left[ 2^{2-n} n \, Q_0 \frac{\Gamma \left( - \frac{n}{2} \right)}{\Gamma \left( \frac{n}{2} \right)} \right]^{1/(2-n)}.
\end{eqnarray}
It is useful to regard the quantity $r_\tau$ as the length scale in the absorbing medium over which the root-mean-square optical depth fluctuation is unity.  The optical depth structure function contains information on the distribution of opacity structure in the absorbing (or emitting) medium, and is a crucial ingredient in determining the image structure of any background source absorbed by such a medium.

\subsubsection{Scintillation characteristics of a source absorbed {\bf or amplified} by a random medium} \label{extendedPower}

As an application of the foregoing formalism, let us consider the practical case of variability from a spectral line in which the power spectrum of the inhomogeneous absorbing medium follows a power law.

Recall that the response of image structure to interstellar scintillation is described in Sect. 2.2, from which we deduce that the mean square of the intensity variations is
\begin{eqnarray}
\langle \Delta S^2 \rangle &=& 8 \pi r_e^2 \lambda^2 \, {\rm SM} \, e^{-2 \langle \tau \rangle + 3 C_\tau(0)/2} \int_{\kappa_{\rm min}} d^2 \bkappa \,  \left\vert V_{\rm src} \left( \frac{\bkappa L}{k} \right) \right\vert^2
 \kappa^{-\beta} \sin^2 \left(\frac{\bkappa^2 L}{2k} \right) \exp \left[ -D_\phi \left( \frac{\bkappa L}{2 k} \right) - \frac{1}{2} D_\tau \left( \frac{\bkappa L}{2 k} \right)  \right]. \label{DeltaSPowIntegral}
\end{eqnarray}

In order to elucidate the properties of spectral line variability from a random absorbing medium, let us consider the simple case of absorption in front of a source of flux density $S_{\rm src}$ and with angular diameter, $\theta_{\rm src}$, larger than or comparable to the refractive angular scale.  Mathematically, this means that the source visibility declines for baselines at $r$ smaller than $(k \theta_{\rm src})^{-1} < r_{\rm ref}^{-1}$.  We assume that the scintillation occurs in the regime of strong scattering where $\sqrt{L/k}> r_{\rm diff}$.  Thus the term $\exp[-D_\phi]$ appearing in Eq. (\ref{DeltaSPowIntegral}) would always cut the integrand off before the sine-squared term begins to vary, so the sine-squared term can be approximated by $\kappa^4 r_{\rm F}^4/4$.   
In practice, 
the upper bound of the integration is determined by whichever of the source or the absorbing medium cuts off the integrand at the lowest wavenumber.  
We further assume that the lower bound of the integration is much smaller than either of the two possible upper bounds.  The resulting variation in the line flux density is 
\begin{eqnarray}
\langle \Delta S^2 \rangle &\approx& S_{\rm ref}^2 \, e^{-2 \langle \tau \rangle + 3 C_\tau(0)/2}  \left\{ \begin{array}{ll}
\left( \frac{\theta_{\rm src}}{\theta_{\rm ref}} \right)^{\beta-6}    , & r_\tau > \frac{1}{k \theta_{\rm src}} \\
\frac{2}{\beta-2} \frac{2-\beta+2n}{(4-\beta+n)} \left( \frac{r_\tau}{r_{\rm diff}} \right)^{6-\beta}, & r_\tau < \frac{1}{k \theta_{\rm src}} \\ 
\end{array}
\right.  , \label{Dstau}
\end{eqnarray}
where 
\begin{eqnarray}
S_{\rm ref}^2 = S_{\rm src}^2 2^{\beta-3} (\beta-2) \frac{\Gamma(\beta/2)}{\Gamma(2-\beta/2)} \left( \frac{r_{\rm diff}}{r_{\rm F}} \right)^{8-2\beta}, \qquad r_{\rm diff} < r_{\rm F}, \label{refpt}
\end{eqnarray} 
is the mean square fluctuation amplitude that a point source of flux density $S_{\rm src}$ undergoing refractive scintillation would exhibit (see Fig. 3).  One clearly sees that the amplitude of the scintillation is enhanced by the additional presence of optical depth variations.  
Even when the source size is large (i.e. when $\theta_{\rm src} > 1/ k r_\tau$), the spectral line variations still differ from those of a continuum source of the same size{\bf ; a} continuum source would exhibit mean square flux density variations of amplitude $S_{\rm ref}^2(\theta_{\rm src}/\theta_{\rm src})^{\beta-6}$, but  the spectral variations are modified by the factor $\exp[-2 \langle \tau \rangle + 3 C_\tau(0)/2]$, which indicates the additional contribution of optical depth variations. The case $r < (k \theta_{\rm src})^{-1}$ is relevant when the optical depth variations occur on exceptionally short scales.  For instance, for a source size $\theta_{\rm src}=1\,$mas this case is relevant when $r_\tau < 3 \times 10^7 \, \lambda$.
We expect this condition to be rarely realised in practice in an absorbing medium, although it may be relevant to some masing systems.

\begin{figure*}
\centering
\includegraphics[width=10cm]{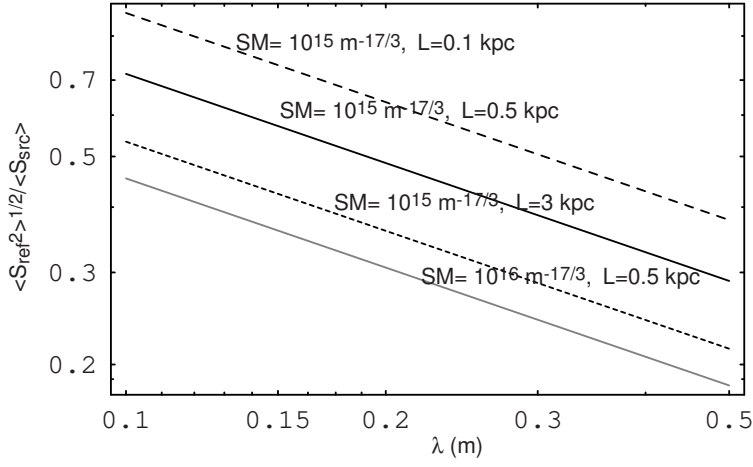} 
\caption{The amplitude of refractive moduluation exhibited by a point source of flux density $S_{\rm src}$ for $\beta=11/3$ and for a number of scattering measures and distances to the scattering medium, as given by Eq. (\ref{refpt}). }
\label{mpref}
\end{figure*}

The root-mean-square flux density deviation is parametrized in terms of the optical depth variance, $\sigma_\tau^2$, and scattering measure, SM, as follows
\begin{eqnarray}
\frac{\langle \Delta S^2 \rangle^{1/2}}{S_{\rm src}} = e^{- \langle \tau \rangle + 3 \sigma_\tau^2 /4} 
\left\{ \begin{array}{ll}
3.28 \times 10^{-21} \, e^{12.52 \beta}
\left( \frac{\lambda}{1\,{\rm m}} \right)^2 \left( \frac{{\rm SM}}{10^{15}\,{\rm m}^{-\beta-2} } \right)^{\frac{1}{2}} \left( \frac{L}{0.5\,{\rm kpc}} \right)^{\frac{\beta}{2}-2} \left( \frac{\theta_{\rm src}}{1\,{\rm mas}} \right)^{\frac{\beta}{2}-3}, & r_\tau > (k \theta_{\rm src})^{-1}, \\
1.31 \times 10^{-22} \, e^{13.11 \beta} \,\sqrt{\frac{\beta-n-2}{(\beta-2)(\beta-n-4)} }\left( \frac{\lambda}{1\,{\rm m}} \right)^{\frac{\beta}{2}-1} \left( \frac{{\rm SM}}{10^{15}\,{\rm m}^{-2-\beta} } \right)^{\frac{1}{2}}  & \null \\
\qquad \qquad  \qquad \qquad \times \left( \frac{L}{0.5\,{\rm kpc}} \right)^{\frac{\beta}{2}-2} \left( \frac{r_\tau}{10^7\,{\rm m}} \right)^{3-\frac{\beta}{2}}, &  r_\tau < (k \theta_{\rm src})^{-1}.
\end{array}  \right.
\end{eqnarray}
For scattering by Kolmogorov turbulence, corresponding to $\beta=11/3$, these scaling relations reduce to
\begin{eqnarray}
\frac{\langle \Delta S^2 \rangle^{1/2}}{S_{\rm src}} = e^{- \langle \tau \rangle + 3 \sigma_\tau^2 /4} 
\left\{ \begin{array}{ll}
0.28  
\left( \frac{\lambda}{1\,{\rm m}} \right)^2 \left( \frac{{\rm SM}}{10^{15}\,{\rm m}^{-17/3} } \right)^{\frac{1}{2}} \left( \frac{L}{0.5\,{\rm kpc}} \right)^{-\frac{1}{6}} \left( \frac{\theta_{\rm src}}{1\,{\rm mas}} \right)^{-\frac{7}{6}}, & r_\tau > (k \theta_{\rm src})^{-1}, \\
0.077 \sqrt{\frac{2n - \frac{5}{3}}{n+\frac{1}{3}} }\left( \frac{\lambda}{1\,{\rm m}} \right)^{\frac{5}{6}} \left( \frac{{\rm SM}}{10^{15}\,{\rm m}^{-17/3} } \right)^{\frac{1}{2}} \left( \frac{L}{0.5\,{\rm kpc}} \right)^{-\frac{1}{6}} \left( \frac{r_\tau}{10^7\,{\rm m}} \right)^{\frac{7}{6}}, &  r_\tau < (k \theta_{\rm src})^{-1}.
\end{array}  \right.
\end{eqnarray}


\section{Application to the variable absorption towards PKS\,1127$-$145}\label{1127}

We now apply the foregoing theory to observations of a variable absorption system.
PKS\,1127$-$145 is a $5-7\,$Jy, $z=1.187$ quasar which exhibits continuum and HI-spectral variability on a time scale of $t\sim 2$\,days (Kanekar \& Chengalur 2001).  The intervening system absorbs 0.5\,Jy, and is located at a redshift $z_{\rm abs}=0.3127$.
After shifting all HI spectra to the same continuum flux level, differences of $\approx 35\,$mJy were found in the strengths of the absorption lines from one observation to the next, showing that the strength of the lines varied relative to the continuum.  The fluctuations are reported to be consistent with variations only in the depths of the components, and not with changes in their positions or widths.  


Kanekar \& Chengalur consider HI variability due to motion of the line of sight towards a hypothetical superluminally ejected subcomponent of the background source.  They argued that the large, rapid continuum variability on inter-day time scales requires $\delta \approx \beta_{\rm app}  \sim 100$.  The time scale of changes in the HI spectrum implies a length scale over which the HI varies significantly in the intervening system.  
Assuming that the temporal variations are due to changes in the line of sight to a superluminal background source, variability on a time scale of $t$ implies substantial changes in the HI structure on the angular scale of (see (\ref{vabs}) in Appendix A) (for $\Omega=1$ and 
$H_0=65$km\,s$^{-1}$\,Mpc$^{-1}$)
\begin{eqnarray}
\theta =  1.7 \, \beta \, \left( \frac{t}{1\,{\rm day}}\right) \, \mu {\rm as}.
\end{eqnarray}
At the redshift of the absorber, $1.7\mu$as corresponds to a transverse scale of $7.7 \times 10^{-4}\,$pc.
For the ejection speed argued by Kanekar \& Chengalur and the observed variability time scale, this implies HI structure on an angular scale of 0.34\,mas.  This angular scale, if attributed to background source motion, is sufficiently small that interstellar scintillation must have at least a significant influence on the HI spectral variability.  


One can also argue that it is not feasible to explain the continuum variability as intrinsic to the source.  
For observations at the frequency of the HI absorption ($\nu_{\rm obs} = 1081\,$MHz), the variability, if intrinsic to the source, implies a brightness temperature
\begin{eqnarray}
T_b = 7.7 \times 10^{20} \left( \frac{S_\nu(\nu_{\rm obs})}{1\,{\rm Jy}} \right) \left( \frac{t_{\rm var}}{1\,{\rm day}}\right)^{-2}~{\rm K},
\end{eqnarray}
The brightness temperature implied by $S_\nu(\nu_{\rm obs})$ variations on a two day time scale requires a Doppler boosting factor in excess of $\delta \sim 700$ for the brightness temperature not to violate the inverse Compton limit in the rest frame of the emitting plasma.  This boosting factor is unacceptably high (Begelman, Rees \& Sikora 1994), and it is not feasible to explain such variability as intrinsic to the source.  Moreover, such variations imply an observed source size 
\begin{eqnarray}
\Omega_{\rm obs} = \Omega \, \delta^{-2} = 7.9 \times 10^{-26} \delta^{-2} \left( \frac{t_{\rm var}}{1\,{\rm day}} \right)^2 .
\end{eqnarray}
Even if the Doppler boosting factor were an order of magnitude less than that implied by the brightness temperature,  the variations require a sub micro-arcsecond angular size.  The source would necessarily undergo scintillation, and would even be expected to exhibit 100\% modulated intensity variations due to diffractive scintillation, which is not observed.  We therefore attribute the continuum variability entirely to interstellar scintillation and consider a model based on scintillation variability alone.  The values derived above are by no means unique.  It should be remembered that the specific combination of source size, absorber size and the degree of absorption associated with the compact cloud are but one of a large number of possible combinations.



In the scintillation model the continuum variations imply a source size
\begin{eqnarray}
\theta_{\rm src} = 35 \left( \frac{t_{\rm var}}{1\,{\rm day}} \right) \left( \frac{v}{30\,{\rm km/s}}\right) \left( \frac{L}{0.5\,{\rm kpc}}\right)^{-1} \, \mu{\rm as},
\end{eqnarray} 
for observations at the absorption wavelength, $\lambda\,0.2773\,$m.  We apply a single cloud absorption model (see Sect. 3.1.2) to observations of this system to explain the amplitude of the variations observed.  Let us suppose $\theta_{\rm src}=0.20$\,mas (e.g. $t_{\rm var}=2\,$days, $v=30$\,km/s and $L=350\,$pc), and take the size of the absorber to be comparable to this at $\theta_{\rm abs}=0.10\,$mas, giving $\theta_{\rm cross}=0.155$\,mas and $\theta_{\rm eff}=0.089\,$mas.  The flux density associated with the compact component in the $5-7$\,Jy source is unknown; we take the flux density of the compact component in the source to be $S_{\rm src}=2.5$\,Jy.  The absorbed flux density is observed to be $\approx 0.5$\,Jy, but if only 2.5\,Jy of the 5\,Jy source is due to absorption against a compact (scintillating) component one has $S_{\rm abs}=0.25$\,Jy.
For a single absorber, the given angular sizes and absorbed flux density imply $A=S_{\rm abs}/S_{\rm src} (\theta_{\rm src}/\theta_{\rm abs})^2=0.4$.  
The variance in spectral line intensity fluctuations that are independent of the continuum fluctuations is (using Eq.\,(\ref{IndepIVar}))
\begin{eqnarray}
\langle \Delta I_{\rm indep}(\nu_{\rm abs})^2 \rangle^{1/2} =  0.070 \left( \frac{C_N^2}{10^{-4}\,{\rm m}^{-20/3} }\right)^{\frac{1}{2}} \left( \frac{L}{0.5\,{\rm kpc}}\right)^{\frac{1}{3}} \, {\rm Jy}.
\end{eqnarray}
For scattering parameters typical of lines of sight off the Galactic plane, $C_N^2 =10^{-4}\,$m$^{-20/3}$ and $L=0.5$\,kpc, the continuum level is expected to exhibit root-mean-square intensity variations of $450$\,mJy and spectral line variations of $170$\,mJy.  The HI line, once the  continuum fluctuations are subtracted, is predicted to exhibit independent root mean square fluctuations of $70$\,mJy.  Thus even the simplest scintillation model is easily capable of explaining the observed 35\,mJy variations in the HI line relative to the continuum in PKS\,1127$-$145.

\section{Discussion}\label{implications}

Scintillation can induce variability in absorption lines relative to the continuum flux density and to other absorption lines.  This is a natural consequence of the fact that images of the source on and off the absorption frequency differ, and that scintillation is sensitive to these differences in image structure.

Scintillation effects are important whenever the source `image' contains significant structure on angular scales comparable to or below the scale of refractive scintillation.  Variations are therefore possible even when the background source is not sufficiently compact to exhibit continuum variability: one only requires absorbing structure to be distributed on a sufficiently small scale.
The angular scale probed by scintillation is typically $\sim 0.1-1$\,mas for observations of HI in the local universe, which corresponds to scales from a few tens of milliparsecs to a few parsecs for the range of distances to extragalactic absorbers.  Studies of HI absorption in our own Galaxy show that HI variations certainly exist on these scales (e.g. Dieter et al. 1976, Diamond et al. 1989), and there is every reason to expect similar variations in other absorbing galaxies.
However, both the size of the background source and the manner in which absorbing material is distributed across it influence the amplitude of any potential spectral variability.

Spectral line variability is expected to be more prevalent in systems absorbed against compact rather than extended sources.  This is because the intensity variations due to structure imposed by clouds distributed randomly against a large background source add together independently, and reduce the amplitude of the modulations relative to the total absorbed flux density.  To see this, consider the variability due to a collection of $N$ clouds absorbing radiation against an extended background source.   The power spectrum of intensity scintillations across the observer's plane is proportional to the power spectrum of the brightness distribution.  The amplitude of the brightness distribution scales only linearly (rather than quadratically) with the number of absorbers if they are distributed randomly and sparsely across the background source.  Thus, for a fixed absorbing flux density the root mean square amplitude of the modulations scales as $N^{-1/2}$.  This is of particular relevance to very large background objects, where the number of foreground absorbers can be correspondingly large. 

When the background source is compact its structure acts like a high-pass filter on the structure of the absorbing material, rendering spectral variability more likely.  In the case of absorption from discrete clouds, this is because the number of independent clouds contributing to the absorption is reduced.  A background source which is itself compact enough to exhibit interstellar scintillation is guaranteed to exhibit spectral line variability relative to the continuum variations provided there is some variation in the {\bf optical} depth across the source.

Scintillation is only capable of modulating the intensities of existing features; it does not alter the frequencies of the spectral lines.  However, this does not preclude variability in the shapes of line profiles.  This is expected if the angular distribution of absorbing material across the background source varies across the line profile, and is particularly relevant to observations in which contributions from individual absorbing features overlap spectrally.  This would be the case if many distinct clouds contribute to the absorption at a single frequency.  In essence, profile variability can occur because the apparent brightness distribution varies within the line profile and causes certain parts of the line profile to be enhanced relative to others; the scattering medium sees different source `images' at different frequencies.  Thus, while scintillation can not shift the absorption frequency, it can alter the line profile by enhancing contributions from one frequency relative to another.

When a spectral feature contains contributions from many absorbing clouds, the outer edges of the line profile are expected to exhibit greater fractional variations relative to the line centre.  The edge of the line profile is more likely to contain very compact structure, due to small amounts of material at extrema of the velocity distribution.  Indeed, such behaviour may have already been observed at the edge of the HI-absorption line profile in PKS\,1127$-$145 by Kanekar \& Chengalur (2001).

Different absorption lines from within the same system are expected to show differing degrees of variability, according to the amount of fine scale structure associated with each absorption line.  Parts of the ISM at different temperatures in an intervening absorbing system are likely to possess different angular distributions.  A colder, more densely clumped medium with more power on smaller angular scales would exhibit more variability.  Such a na\"{\i}ve model would predict narrow lines to exhibit more variability relative to the broader lines from slightly warmer parts of the medium.

The response of scintillation to a power law spectrum of absorption (emission) fluctuations is determined by the variance and scale length of the optical depth fluctuations in the medium.  For a background absorbing source larger than or comparable to the refractive scale, the presence of the optical depth fluctuations enhances the mean square amplitude fluctuations by a factor $\exp(3 \langle \tau^2 \rangle/2)$.  This enhancement is due to the extra structure imprinted on the background source by the absorbing medium.

Scintillation-induced spectral variability is more important for absorbers at high redshifts.  Absorbers of a fixed spatial scale appear smaller at high redshifts.  The apparent angular diameter of structure of a fixed linear scale, $s_0$, decreases with redshift as $s_0/D_A(z)$ for angular diameter distance $D_A(z)$.  As the angular diameter distance increases until at least $z \ga 1$ (depending on the cosmology assumed), so does the apparent source size decrease.
A second effect also enhances spectral variability at high redshift.  The angular scale probed by the refractive scintillations increases with wavelength as $\lambda^{2.2}$.  Scintillation effects are most important when this scale exceeds the angular scale of the absorption structure.  It is easier for absorption which occurs at higher redshifts, and thus at longer apparent wavelengths, to satisfy this criterion.

Several other possible causes of spectral line variability have been suggested.  Lewis \& Ibata (2003) discuss the contribution of gravitational lensing to absorption line variability.  Like scintillation, variability due to gravitational lensing requires the presence of small-scale structure in the absorbing material.  However, the presence of lensing material is problematic: it is not clear whether the lensing optical depth in intervening systems is sufficiently large to reproduce the continuous sort of variability that is reported.  On the other hand, if lensing effects are important then it is likely that scintillation variability will also contribute strongly to the variability, as both mechanisms depend on the presence of small scale structure in the absorbing medium.  

Spatial variations in the absorbing structure coupled with motions in the background source can also cause line variability.  Both Kanekar \& Chengalur (2001) and the present paper discuss the possibility of source motions giving rise to HI variability in PKS\,1127$-$145, while Briggs (1983) considers this in relation to the HI variations observed in AO 235$+$164.  
For the model considered by Briggs (1983), the optical depth variations are not independent between spectral lines. Measurements of any two lines are sufficient to predict the variations of the other lines observed. Application of this model to AO 235$+$164 shows that background source motions can reproduce the spectral variability observed.  However, we note that the presence of correlations between spectral line variations does not preclude scintillation as a viable explanation, as scintillation causes correlated variability between sets of spectral lines whenever there is a correlation in the spatial distribution of the absorbing material.  Indeed, if the structure were all located within a small enough angle (i.e. within $\theta_{\rm ref}$), the variability would be completely correlated.





\section{Conclusions}\label{conclusions}
We have presented a theory for the effect of interstellar scintillation on the variability of HI absorption lines from extended (i.e. non point-like) extragalactic sources.  The model predicts modulations of the line strengths relative to the continuum and relative to other lines.  It does not permit shifts in the frequency of the lines.   However, apparent shifts in the line frequency can occur if the angular distribution of the absorbing structure differs with frequency within the line.  

Scintillation is most sensitive to structure on scales less than the refractive scale of the scintillation, typically $\sim 1\,$mas.  Line variability arises when intervening absorption imposes structure in the source image that varies significantly with frequency on this angular scale.

The theory presented here applies equally to any  absorption or emission line in the radio band provided that its distribution is sufficiently compact for scintillation to be important.  In this respect, the foregoing theory is also useful in the interpretation of fast maser variability (Greenhill et al. 1997).

Other possible explanations of spectral line variability include gravitational lensing effects and motions in the position of the background source relative to the absorbing material.  However, the presence of small scale structure in the absorbing material in both cases means that scintillation effects are likely to play an important, if not dominant role in the variability.  Scintillation is a viable explanation for the cases of spectral variability hitherto observed at radio wavelengths, particularly for observations of the HI line.

As an application of the theory, we have examined the HI variability reported along the line of sight to PKS\,1127$-$47, in which spectral lines vary relative to one another and to the continuum.  A simple model accounts for all the observed inter-day spectral variability.  At this frequency, scintillation can account for variations on time scales of days to weeks.  We also argue that scintillation must be important in this source even as an explanation of the continuum variability.  


\begin{acknowledgements}
I thank Nissim Kanekar for many useful discussions and for reading the manuscript critically and Avinash Desphande for reviewing the manuscript and pointing out a conceptual error in the original manuscript.
\end{acknowledgements}

\appendix

\section{Time scales for superluminal motion of a background source relative to an intervening absorbing medium}
Often the emergence of a new compact background component is associated with the relativistic ejection of material from a background quasar.  The apparent transverse velocity of the component often exceeds $c$, and one expects the point at which the line of sight intersects an intervening absorption system to change with time.  Here we compute the time scale of variability due to such motion.

Suppose an observer at Earth measures a change in angle $\mu$ in the direction to the background source over a time interval.  The change in angle corresponds to a distance $x_{\rm src}=\mu D_A(z_{\rm src})$ at the source, where $D_A(z_{\rm src})$ is the angular diameter distance, given by
\begin{eqnarray}
D_A(z) = \frac{2c}{H_0} \left( \frac{\Omega z + (\Omega-2) (\sqrt{1+\Omega z}-1)}{\Omega^2 (1+z)^2} \right).
\end{eqnarray}
The rate of change of the angle $\mu$ (i.e. the proper motion) measured by an observer on Earth can be related to the velocity
\begin{eqnarray}
\frac{d \mu}{dt_{\rm E}}  &=& \frac{d}{dt_{\rm E}} \frac{x_{\rm src}}{D_A(z_{\rm src})} \nonumber \\
&=& \frac{dt_{\rm src}}{dt_{\rm E}} \frac{d}{dt_{\rm src}} \left( \frac{x_{\rm src}}{D_A(z_{\rm src})} \right) \nonumber \\
&=& \frac{v_{\rm src}}{(1+z_{\rm src}) \, D_A(z_{\rm src})}.   \label{muDot}
\end{eqnarray}
We have used the fact that frequencies in the source frame and Earth frame are related by $\nu_{\rm E}/\nu_{\rm src} = dt_{\rm src}/dt_{\rm E} =(1+z)^{-1}$;  
an event of duration $\Delta t_{\rm E}$ measured on Earth corresponds to a duration $\Delta t_{\rm E}/(1+z)$ in the frame of the emitting system.  A source expanding at $v_{\rm src}=\beta c$ as measured at the redshift of the source has a proper motion 
\begin{eqnarray}
\dot{\mu} = \frac{\beta  c}{ D_A (1+z_{\rm src})}.
\end{eqnarray}
The proper motion is determined by the apparent speed of the material, which may exceed $c$ by a factor of tens, and by the redshift.  We can also use Eq. (\ref{muDot}) to deduce the proper velocity, $v_{\rm abs}$, at which the line of sight moves across an intervening HI absorbing screen, located at redshift $z_{\rm abs}$
\begin{eqnarray}
v_{\rm abs} = \dot{\mu} D_A(z_{\rm abs}) (1+z_{\rm abs}).
\end{eqnarray}
The speed at which the line of sight moves across the HI absorbing screen is thus
\begin{eqnarray}
v_{\rm abs} = \beta_{\rm app} \, c \, \frac{D_A(z_{\rm abs})}{D_A(z_{\rm src})} \left( \frac{1+z_{\rm abs}}{1+z_{\rm src}} \right). \label{vabs}
\end{eqnarray}

\section{Average power spectrum for a large number of absorbers in front of a compact source}\label{ManyAbsorbers}
If there is a large number of absorbers covering the background source it is appropriate to average over their positions.  The average over $\Delta \btheta_i$ is performed by assuming they are randomly distributed in space out to some angular distance, $\theta_{\rm max}$ with a probability density function
\begin{eqnarray}
p(\Delta \btheta) = \left\{ \begin{array}{ll} 
(2 \theta_{\rm max})^{-2} , & |\Delta \theta_{x}| < \theta_{\rm max}, \, |\Delta \theta_{y}| < \theta_{\rm max}, \\
0, & {\rm otherwise} \\
\end{array} \right. .
\end{eqnarray}
The average over positions is performed by integrating over the probability distribution 
\begin{eqnarray}
&\null&  \int_{-\theta_{\rm max}}^{\theta_{\rm max}} \frac{d\Delta \theta_x  d\Delta \theta_y}{4 \theta_{\rm max}^2} \, 
 \exp \left[ i (a_x \Delta \theta_x + a_y \Delta \theta_y) - b (\Delta \theta_x^2 + \Delta \theta_y^2)  \right] \nonumber \\
&\null&  \qquad =  \frac{\pi}{16 \, b \,\theta_{\rm max}^2} \exp \left( -\frac{a_x^2 + a_y^2}{4 b} \right) \nonumber 
\left[ 
{\rm erf} \left( - \frac{2 b \theta_{\rm max} + i a_x}{2 \sqrt{b}} \right) -
{\rm erf} \left( \frac{2 b \theta_{\rm max} - i a_x}{2 \sqrt{b}} \right) 
\right] \nonumber \\
&\null& \qquad \qquad \qquad \qquad \times
\left[ 
{\rm erf} \left( - \frac{2 b \theta_{\rm max} + i a_y}{2 \sqrt{b}} \right) -
{\rm erf} \left( \frac{2 b \theta_{\rm max} - i a_y}{2 \sqrt{b}} \right)
\right],
\end{eqnarray}
where 
\begin{eqnarray}
(a_x,a_y) &=& \frac{L \theta_{\rm src}^2}{\theta_{\rm src}^2 + \theta_1^2} (\kappa_x,\kappa_y), \\
b &=& \frac{1}{\theta_{\rm src}^2 + \theta_1^2}. 
\end{eqnarray}

\section{The average mean square visibility due to a random absorbing or emitting medium} \label{TauDeriv}

The wavefield incident upon a point ${\bf r}$ on a plane at a distance $D$ from a thin screen of absorbing material with phase variations $\phi$ can be written in terms of the Fresnel-Kirchoff integral (see, e.g., Born \& Wolf 1965, Goodman \& Narayan 1989)
\begin{eqnarray}
u({\bf r}) = \frac{e^{ik D}}{2 \pi i r_{\rm F}^2 } \int d^2 {\bf x} \, u_{\rm src}({\bf x}) \exp \left[i \frac{({\bf x} -{\bf r}) ^2}{2 r_{\rm F}^2} \right] e^{ i \phi({\bf x})}  , \label{TauFK}
\end{eqnarray}
where $u_{\rm src}({\bf x}')$ is the wavefield of the background source incident upon the absorbing medium and the Fresnel scale is given by $r_{\rm F} = \sqrt{D/k}$.  The background source is assumed to be located at infinity, so that its wavefronts incident upon the absorbing plane are planar.  

Here $\phi$ is regarded as a complex quantity.  Since we are interested only in opacity variations, we henceforth set Re$(\phi)=0$ and consider only the imaginary contribution to the phase, Im($\phi) \equiv \eta = \tau/2$.  With this specification, Eq. (\ref{TauFK}) is a solution to the wave equation for a thin screen of absorbing material in the absence of refraction.  

Equation (\ref{TauFK}) incorporates wave effects that are normally neglected when considering the effect of optical depth variations.  For instance, if one points a telescope toward a particular point in an absorbing medium, one expects to measure an optical depth associated with that particular point in the medium.  However, although the optical depth $\tau$ is a well-defined quantity, Eq. (\ref{TauFK}) implies that the wavefield one measures contains the contributions $\tau$ averaged over some region in the absorbing material, with the weighting function given by the first exponential term appearing in the integrand.  This effect is important when the optical depth variations occur on a scale that is small compared to the length scale $r_{\rm F}=\sqrt{D/k}$.   Equation (\ref{TauFK}) reduces to the familiar definition of optical depth $u({\bf x}) = u_{\rm src}({\bf x}) e^{-\tau({\bf x})/2}$ in the limit of geometric optics, which corresponds to $\lambda \rightarrow 0$.  However, we choose to work within the framework of physical optics in the present case because its generality allows one to consider the effect of absorbing structures that may be small compared to $r_{\rm F}$. This case may be relevant for absorption by some extragalactic systems where $D$ is large and the medium may be sufficiently inhomogeneous.

The mean square visibility of radiation from a distant absorbed background source which has propagated through the random absorbing medium is
\begin{eqnarray}
\langle V_{\rm abs}({\bf r}') V_{\rm abs}^*({\bf r}') \rangle &=&\langle  u({\bf r}+{\bf r}'/2) u^*({\bf r}-{\bf r}'/2) u({\bf r}-{\bf r}'/2) u^*({\bf r}+{\bf r}'/2) \rangle \nonumber \\
&=& \frac{1}{(2 \pi r_{\rm F}^2)^4}  \int d^2 {\bf x}_1 d^2{\bf x}_2 d^2{\bf x}_3 d^2{\bf x}_4 \, \left \langle u_{\rm src}({\bf x}_1)  u_{\rm src}^*({\bf x}_2) u_{\rm src}({\bf x}_3) u_{\rm src}^*({\bf x}_4) \right\rangle \left\langle  e^{- \eta({\bf x}_1) - \eta({\bf x}_2)-  \eta({\bf x}_3) - \eta({\bf x}_4) }  \right\rangle \nonumber \\ 
&\null& \quad  \qquad \times \exp \left[  i \frac{ 
({\bf x}_1-{\bf r}-{\bf r}'/2)^2 - ({\bf x}_2-{\bf r}+{\bf r}'/2)^2+ ({\bf x}_3-{\bf r}+{\bf r}'/2)^2 - ({\bf x}_4-{\bf r}-{\bf r}'/2)^2 }{2 r_{\rm F}^2 } \right].
\end{eqnarray}
For a spatially incoherent background source, the average over the source wavefield appearing in the integral reduces to a product of  visibilities, $V({\bf x}_1-{\bf x}_2) V^*({\bf x}_3-{\bf x}_4)$.  We use the result $\langle \exp[-x] \rangle = \exp[-\langle x \rangle + \langle x^2 \rangle/2]$ for a normally distributed variate $x$ to reduce the average over opacity variations: 
\begin{eqnarray}
\left\langle  e^{- \eta({\bf x}_1) - \eta({\bf x}_2)-  \eta({\bf x}_3) - \eta({\bf x}_4) }  \right\rangle 
&=& \exp \Big\{-4 \langle \eta \rangle + \frac{1}{2} \big[ 12 C_\eta (0) -D_\eta ({\bf x}_1-{\bf x}_2) - D_\eta ({\bf x}_1-{\bf x}_3) - D_\eta ({\bf x}_1-{\bf x}_4)  \nonumber \\
&\null&  \qquad \qquad - D_\eta ({\bf x}_2-{\bf x}_3) -D_\eta ({\bf x}_2-{\bf x}_4) - D_\eta ({\bf x}_3-{\bf x}_4)
\big] \Big\} , 
\end{eqnarray} 
where $C_\eta ({\bf r}) = \langle [\eta({\bf r}+{\bf r}') - \langle \eta \rangle]\,[\eta({\bf r}') - \langle \eta \rangle] \rangle$ is the autocovariance of opacity variations and $D_\eta ({\bf r}) = 2 [C_\eta(0) - C_\eta ({\bf r})]$ is its corresponding structure function.

Making the change of variables
\begin{eqnarray}
{\bf R} &=& \frac{1}{4}[{\bf x}_1+{\bf x}_2 + {\bf x}_3 + {\bf x}_4] \nonumber \\
{\bf r}_1&=& \frac{1}{2} [{\bf x}_1 + {\bf x}_2 - {\bf x}_3 - {\bf x}_4] \nonumber \\
{\bf r}_2&=& \frac{1}{2}  [{\bf x}_1 - {\bf x}_2 - {\bf x}_3 + {\bf x}_4] \nonumber \\
{\bf \rho} &=& {\bf x}_1-{\bf x}_2 + {\bf x}_3 - {\bf x}_4,
\end{eqnarray}
the mean square visibility reduces to 
\begin{eqnarray}
&\null&\langle  u({\bf r}+{\bf r}'/2) u^*({\bf r}-{\bf r}'/2) u({\bf r}+{\bf r}'/2) u^*({\bf r}-{\bf r}'/2) \rangle \nonumber \\
&\null& \qquad = \frac{1}{(2 \pi r_{\rm F}^2)^4}  \int d^2 {\bf R} \, d^2{\bf \rho} \, d^2{\bf r}_1 \, d^2{\bf r}_2 \, \exp \left[ i \frac{  {\bf r}_1 \cdot ({\bf r}_2-{\bf r}') + {\bf \rho} \cdot {\bf R} - {\bf \rho}\cdot {\bf r} }{r_{\rm F}^2}\right] 
V_{\rm src}({\bf r}_2 + {\bf \rho}/2) V_{\rm src}({\bf \rho}/2 - {\bf r}_2) \nonumber \\
&\null& \qquad \qquad 
\exp \Big\{-4 \langle \eta \rangle + \frac{1}{2} \big[ 12 C_\eta (0) -D_\eta ({\bf r}_2 + {\bf \rho}/2) - 
D_\eta ({\bf r}_1+{\bf r}_2) - D_\eta ({\bf r}_1+{\bf \rho}/2)  - D_\eta ({\bf r}_1-{\bf \rho}/2)  \nonumber \\
&\null&  \qquad \qquad  \qquad \qquad  -D_\eta ({\bf r}_1-{\bf r}_2) - D_\eta ({\bf \rho}/2 - {\bf r}_2)
\big] \Big\} .
\end{eqnarray}
The integral over ${\bf R}$ yields a factor $\delta({\bf \rho}) (2 \pi r_{\rm F}^2)^2$, and the subsequent integration over ${\bf \rho}$ eliminates ${\bf r}$, yielding,
\begin{eqnarray}
&\null&\langle  u({\bf r}+{\bf r}'/2) u^*({\bf r}-{\bf r}'/2) u({\bf r}+{\bf r}'/2) u^*({\bf r}-{\bf r}'/2) \rangle \nonumber \\
&\null& \qquad = \frac{1}{(2 \pi r_{\rm F}^2)^2}  \int  d^2{\bf r}_1 \, d^2{\bf r}_2 \, \exp \left[ i \frac{ ({\bf r}_2-{\bf r}') \cdot {\bf r}_1 }{r_{\rm F}^2}\right] 
V_{\rm src}({\bf r}_2) V_{\rm src}^*({\bf r}_2) \nonumber \\
&\null& \qquad \qquad 
\exp \Big\{-4 \langle \eta \rangle + \frac{1}{2} \big[ 12 C_\eta (0) - 2 D_\eta ({\bf r}_1) - 2 D_\eta ({\bf r}_2) 
- D_\eta ({\bf r}_1+{\bf r}_2) -D_\eta ({\bf r}_1-{\bf r}_2) 
\big] \Big\} . \label{PhysOptVisSqr}
\end{eqnarray}

In the geometric optics limit wave effects are removed by taking the limit $\lambda \rightarrow 0$, which corresponds to the limit $r_{\rm F} \rightarrow 0$.  In this case one can show, either from Eq. (\ref{TauFK}) or directly from Eq. (\ref{PhysOptVisSqr}), that the mean square visibility is
\begin{eqnarray}
\langle | V_{\rm abs}({\bf r}') | ^2 \rangle = | V_{\rm src}({\bf r}) |^2 \exp \left[ -4 \langle \eta \rangle + 6 C_\eta(0) - 2 D_\eta ({\bf r}') \right].
\end{eqnarray}



\begin{thebibliography}{}

\bibitem[1995]{Armstrong} Armstrong, J.W., Rickett, B.J. \& Spangler, S.J. 1995, \apj, 443, 209
\bibitem[1999]{Braun} Braun, R. 1999, in ``Interstellar Turbulence'', Proceedings of the 2nd Guillermo Haro Conference, eds. J. Franco \& A. Carraminana, Cambridge University Press, 12
\bibitem[1994]{Begelman} Begelman, M.C., Rees, M.J. \& Sikora, M. 1994, \apj, 429, L57 
\bibitem[2003]{Bignalletal} Bignall, H.E., Jauncey, D.L., Lovell, J.E.J., Tzioumis, A.K.,
Kedziora-Chudczer, L., Macquart, J.-P., Tingay, S.J., Rayner, D.P. \& Clay,
R.W. 2003, \apj, 585, 683
\bibitem[1965]{Born} Born, M. \& Wolf, E., Principles of Optics, Oxford: Permagon Press,
1965
\bibitem[1983]{Briggs} Briggs, F.H. 1983, \apj, 274, 86
\bibitem[1986]{Codona} Codona, J.L. \& Frehlich, R.G. 1987, Radio Sci., 22, 469
\bibitem[1987]{Colesetal} Coles, W.A., Frehlich, R.G. Rickett, B.J. \& Codona, J.L. 1987, \apj, 315, 666
\bibitem[1976]{Deiter} Deiter, N.H., Welch, W.J. \& Romney, J.D. 1976, \apj, 206, L113
\bibitem[1999]{Dennett-Thorpe99} Dennett-Thorpe, J. \& de Bruyn, A.G., \apj, 529, L65 
\bibitem[2003]{Dennett-Thorpe03} Dennett-Thorpe, J. \& de Bruyn, A.G., \aa, 404, 113
\bibitem[1981]{DennisonCondon} Dennison, B. \& Condon, J.J. 1981, \apj, 246, 91
\bibitem[2000]{Desphande} Deshpande, A.A., Dwarakanath, K.S. \& Goss, W.M. 2000, \apj, 543, 227
\bibitem[1989]{Diamondetal} Diamond, P.J., Goss, W.M., Romney, J.D., Booth, R.S., Kalberla, P.M.W. \& Mebold, U. 1989, \apj, 347, 302
\bibitem[1993]{Green} Green, D.A. 2000 \mnras, 262, 327
\bibitem[1989]{Goodman} Goodman, J. \& Narayan, R. 1989, \mnras, 238, 995
\bibitem[1997]{Greenhilletal} Greenhill, L.J., Ellingsen, S.P., Norris, R.P., Gough, R.G., Sinclair, M.W., Moran, J.M., Mushotzky, R. 1997, \apj, 474, L103
\bibitem[2001]{Gwinn} Gwinn, C.R. 2001, \apj, 561, 815
\bibitem[2001]{KanekarChengalur} Kanekar, N. \& Chengalur, J.N. 2001, \mnras, 325, 631
\bibitem[1997]{Kedziora-Chudczer} Kedziora-Chudczer, L., Jauncey, D.L., Wieringa, M.H., Walker, M.A.,
Nicolson, G.D., Reynolds, J.E. \& Tzioumis, A.K. 1997, \apj, 490, L9  
\bibitem[2001]{Kedziora-Chudczer01} Kedziora-Chudczer, L., Macquart, J.-P. \& Jauncey, D.L. in ``Galaxies and their Constituents at the Highest Angular Resolutions'', Proceedings of IAU Symposium 205, ed. R. T. Schilizzi, 2001, 90
\bibitem[2000]{LazarianPogosyan} Lazarian, A. \& Pogosyan, D. 2000, ApJ, 537, 720
\bibitem[1966]{LittleHewish} Little, L.T. \& Hewish, A 1966, \mnras, 134, 221
\bibitem[1962]{Mercier} Mercier, R.P. 1962, Proc. Cambridge Phil. Soc., 58, 382
\bibitem[1965]{Papoulis} Papoulis, A. Probability, random variables, and stochastic processes, New York : McGraw Hill Book Company, 1965
\bibitem[1975]{Prokhorovetal} Prokhorov, A.M., Bunkin, F.V., Gochelashvity, K.S. \& Shishov, V.I. 1975, Proc. IEEE, 63, 790
\bibitem[1977]{Rickett77} Rickett, B.J., 1977, \araa, 15, 479
\bibitem[2001]{Rickett} Rickett, B.J. 2001, Ap\&SS, 278, 129
\bibitem[1975]{Rumsey} Rumsey, V.H. 1975, Radio Sci., 10, 107
\bibitem[1967]{Salpeter} Salpeter, E.E. 1967, \apj, 147, 433
\bibitem[1993]{Taylor} Taylor, \& Cordes, J.M., 1993, \apj, 411, 674
\bibitem[1998]{Walker} Walker, M.A. 1998, 294, 307
\bibitem[1982]{Wolfeetal} Wolfe, A.M., Davis, M.M. \& Briggs, F.H. 1982, \apj, 259, 495


%

\end{thebibliography}
\end{document}